\documentstyle[12pt,preprint,aps,pra]{revtex}
\begin{document}
\draft
\title{STABILITY OF THE HYDROGEN\\
AND HYDROGEN-LIKE MOLECULES}
\author{Jean-Marc RICHARD}
\address{
Institut des Sciences Nucl\'eaires\\
53, avenue des Martyrs, 38026 Grenoble, France\\
and\\
European Centre for Theoretical Studies\\
in Nuclear Physics and Related Areas (ECT*)\\
strada delle Tabarelle 286, Villazzano (Trento),
Italy}
\date{\today}
\maketitle
\begin{abstract}
We present a simple proof
of the stability of the hydrogen molecule $(M^+M^+m^-m^-)$.
It does not rely on the proton-to-electron mass ratio
$M/m$ being very large, and actually holds for arbitrary values
of $M/m$. Some asymmetric molecules of the type
$(m_1^+m_2^+m_3^-m_4^-)$ are
also stable. Possible applications to  molecules
containing antiparticles and to
exotic hadrons in the quark model are briefly outlined.
\end{abstract}
\pacs{13.35.Cs}
\section{Introduction}
\label{Intro}
The study of the hydrogen molecule is usually carried out in the
framework of the Born--Oppenheimer approximation \cite{BasicQM}.
The effective potential between the two protons exhibits a deep
pocket of attraction, and there is no doubt that stability survives
quantum fluctuations and other corrections due to  proton mass
being finite. A simple and rigourous derivation of the stability
is, however, desirable, as well as a systematic study of other
4-unit-charge systems.

A proof of the stability of the hydrogen molecule was briefly
outlined in a recent communication \cite{BadHonnef}. It does not
rely on the proton mass $M$ being very large as compared to the
electron mass $m$. It is actually based on a result
obtained for the $M=m$ case: Hylleraas and Ore \cite{Ore} have
shown in 1947 that the positronium molecule $(e^+e^+e^-e^-)$ is
stable against dissociation into two positronium atoms. The
stability of the hydrogen molecule, and more generally of any
$(M^+M^+m^-m^-)$ configuration,  can be proved using a rescaled
version of the wave function  of the
positronium molecule.

If one aims at mathematical
rigour, one can object that in the analysis of
Refs.\ \cite{BadHonnef,Ore}, it is
implicitly assumed that the lowest threshold consists
of two neutral atoms. It is also desirable to make connection with
the usual approach to hydrogen binding, based on the
Born--Oppenheimer limit $M/m\rightarrow\infty$.
In a recent letter\cite{Froehlich}, the question of the thresholds is
seriously addressed, and an alternative variational wave function
was proposed, which proves stability for $M\ge 0.144m$, and
reduces to the standard Heitler--London wave function in the limit
$M/m\rightarrow\infty$. A more detailed account of this work on the
thresholds and on the stability for $M\gg m$ will be published
elsewhere \cite{Seifert}.

In the present paper, we explain how stability
in the equal-mass case $(e^+e^+e^-e^-)$ implies stability for all
hydrogen-like configurations $(M^+M^+m^-m^-)$. Moreover, from the
calculated  binding energy of $(e^+e^+e^-e^-)$ with respect to its
threshold $(e^+e^-)+(e^+e^-)$, we predict  a
minimal extension of the stability region toward even more asymmetric
configurations of the type $(m_1^+m_2^+m_3^-m_4^-)$.

Perhaps the stability of
the Coulombic molecule $(p\Sigma^+\Omega^-\Xi^-)$  with baryon number
$B=4$ and strangeness $S=4$ will
be thought as being of marginal interest.  (For the zoology of charged
leptons, mesons, and baryons, we refer to the {\sl Review of Particles
Properties} \cite{PDG}.)  There  is  however a very general and
fundamental concern about understanding why matter
sometimes shows up in large compounds and sometimes splits
into small clusters. We hope to better penetrate the mechanism of
molecular binding by studying how it is sensitive to the masses of
the constituents which are involved.
Moreover, our new approach to hydrogen binding
allows for some generalisations. As briefly outlined in
Sec.~\ref{Larger-molec},  exotic molecules involving  a
mixture of matter and antimatter might well be stable,
as long as annihilation is  neglected.  The stability problem also
exists in other fields. In quark physics, only the minimal
configurations have  been seen so far, namely mesons made out of a
quark and an antiquark, and baryons consisting of three
quarks. We shall mention in Sec.~\ref{sec-tetraquark} that the patterns
observed in molecular physics might provide some guidance
in quark physics, or at least in simple quark models, to guess
which flavour configurations are the most likely to host stable
multiquarks.

\section{ The stability problem}
\label{Stability}

Let us consider four elementary particles, with masses $m_i$,
and unit electric
charges $q_i=(+1,+1,-1,-1)$. There are obvious symmetries:
an overall charge conjugaison, as well as $(1\leftrightarrow2)$
and $(3\leftrightarrow4)$ exchanges, leave the problem
of stability invariant. So one can assume without loss of
generality that
\begin{equation}
\label{mass-ordering}
m_1\ge m_2,\qquad m_3\ge m_4,
\end{equation}
and that  the positive charges are
on the average heavier than the negative ones,
if there is any overall asymmetry.

In most cases, instability manifests itself by dissociation
into two neutral atoms. For some values of the
mass ratios,  the lowest threshold consists of a 3-body ion
and a charge, for instance $(m_1m_2m_3)^++m_4^-$: we know that
$(m_1^+m_2^+m_3^-)$ is stable in a wide range of
values of $m_2/m_1$ and $m_3/m_1$ \cite{ArmB,MRW};
in particular, every
ion with $m_1=m_2$ is stable \cite{Hill}; moreover, if $m_1$, $m_2$,
and $m_3$ are large, the binding energy of  $(m_1m_2m_3)^+$ might
well exceed (in absolute value) that of every other threshold.
However, a molecule $(m_1^+m_2^+m_3^-m_4^-)$ would hardly decay
into $(m_1m_2m_3)^++m_4^-$, since the long-range  Coulomb
attraction would keep the ion and the charge together. In other
words, if the lowest threshold is made of a 3-body ion and a
charge, the stability of the 4-body molecule is guarantied.
We shall come back to this point in Sect.~\ref{Very-asymmetric}.

The difficult task is thus to compare the 4-body ground state
to the thresholds consisting of two neutral atoms. Out of the two
possible arrangements, $(1,3)+(2,4)$ and
$(1,4)+(2,3)$, the former is more stable, if
one assumes the mass ordering (\ref{mass-ordering}). This can be
seen directly from  the Bohr formula for the
2-body energy,  $E_2(m_i^+,m_j^-)=
-1/(2\alpha_{ij})$, with
$\alpha_{ij}=m_i^{-1}+m_j^{-1}$ being
the inverse reduced mass. This is
in fact a rather general result \cite{BerMar}. If $v(r)$ is
an  universal potential (independent of the
masses), the  reduced 2-body
Hamiltonian $h=\alpha\vec{\rm p}^2/2+v(r)$ depends
on the inverse reduced mass $\alpha$ linearly.
Then the ground state energy $E_2(\alpha)$ is a
concave function of $\alpha$~\cite{Thir},
and if (\ref{mass-ordering}) is satisfied,
$E_2(\alpha_{13})+E_2(\alpha_{24})\le
E_2(\alpha_{14})+E_2(\alpha_{23})$.

Some states lying above the threshold are metastable, since their
decay involves painful rearrangements and tunnelings. We shall not
consider them.

One should also mention unnatural parity states. Imagine a molecule
carrying orbital angular momentum and parity $J^P=1^+$, besides the
intrinsic spins and parities of its constituents.
It cannot decay into two ground-state
atoms $j^p=0^+$ separated by an orbital momentum $\ell=1$. Its
actual threshold is made of a ground state (1S) and an
orbital excitation (2P). We shall not consider this problem
further, but simply mention that some of the general
methods and results presented in this  article can be
extended to unnatural-parity configurations.

\section{The equal-mass case}
\label{Positronium}
The stability of the positronium molecule was
shown in 1947  by Hylleraas and Ore \cite{PDG}, who
used an elegant variational method.
They first got rid of the scale by noticing that if
$\Psi(\vec{\rm r}_i)$ is a trial wave function, with
norm and expectation values of the kinetic and potential
energies written as
\begin{equation}
\label{n-t-v}
n=\langle\Psi|\Psi\rangle,\quad
t=\langle\Psi|T|\Psi\rangle,\quad
v=\langle\Psi|V|\Psi\rangle,
\end{equation}
then the best rescaling  of the type
$\phi=\Psi(\vec{\rm r}_i/\lambda)$ yields a minimum
\begin{equation}
\label{virial}
\widetilde{E}=-{v^2\over4tn},
 \end{equation}
which corresponds to $\langle\phi|T|\phi\rangle
=-\langle\phi|V|\phi\rangle/2$, i.e.\ the same sharing of the
kinetic and potential energies as for the exact
solution. This extension of the virial theorem to
variational approximations is well known\cite{Fock}.

The frozen-scale wave function of Hylleraas and Ore contains
a single parameter:
\begin{equation}
\label{Hylleraas-wf}
\Psi=\exp-{1\over2}(r_{13}+r_{14}+r_{23}+r_{24})
\cosh{\beta\over2}(r_{13}-r_{14}-r_{23}+r_{24}).
\end{equation}
Explicit integration leads to (a misprint in Ref.\ \cite{Ore}
is corrected below)
\begin{mathletters}
\begin{eqnarray}
\label{Hylleraas-ev}
n&=&{33\over16}+{33-22\beta^2+5\beta^4\over16(1-\beta^2)^3}\\
t&=&{21\over8}-{3\beta^2\over2}
  +{21-6\beta^2+\beta^4\over8(1-\beta^2)^3}\\
v&=&{19\over6}+{21-18\beta^2+5\beta^4\over4(1-\beta^2)^3}\nonumber\\
& &-
{1\over(1-\beta^2)^2}\left[1-{5\beta^2\over8}-{1\over4\beta^4}
+{7\over8\beta^2}+{(1-\beta^2)^4\over4\beta^6}\ln
{1\over1-\beta^2}\right],
\end{eqnarray}
\end{mathletters}
to be inserted in (\ref{n-t-v}) and (\ref{virial}), leading to a
minimum $\widetilde{E}=-0.5042$ near
$\beta^2=0.48$.

It is rather easy to generalise the calculation of Ref.\ \cite{Ore}
to a trial wave function of the type
\begin{equation}
\label{Hylleraas-wf-gen}
\Psi=\sum_i
c_i\exp-a_i(r_{13}+r_{14}+r_{23}+r_{24})
\cosh b_i(r_{13}-r_{14}-r_{23}+r_{24}),
\end{equation}
but one does not gain much \cite{Ore2}. As analysed for instance by Ho
\cite{Ho} and by the authors he quotes, some explicit $r_{12}$ and
$r_{34}$ dependence is needed in the wave function to improve the
accuracy. The last variational calculation \cite{Kozl}
gives an energy one can express as
\begin{equation}
\label{Posit-x}
E({\rm e}^+{\rm e}^+{\rm e}^-{\rm e}^-)=
-(1+x_1)/2,\qquad x_1=0.03196,
\end{equation}
where $x_1$ represents the fraction of additional binding
with respect to the threshold, which is $E_{\rm th}=-1/2$ in our
units.

\section{From the positronium to the hydrogen molecule}
\label{Hydrogen}
Once the stability of $(e^+e^+e^-e^-)$ is established, it is extremely
simple to  derive that of every $(M^+M^+m^-m^-)$ configuration
\cite{BadHonnef,Froehlich}.
Let us, indeed, rewrite the Hamiltonian as
\begin{mathletters}
\begin{eqnarray}
\label{split-H}
H&=&H_{\rm S}+H_{\rm A}\\
H_{\rm S}&=&\left({1\over4M}+{1\over4m}\right)
\left(\vec{\rm p}_1^2+\vec{\rm p}_2^2+\vec{\rm p}_3^2
+\vec{\rm p}_4^2\right)+V\\
H_{\rm A}&=&\left({1\over4M}-{1\over4m}\right)
\left(\vec{\rm p}_1^2+\vec{\rm p}_2^2-\vec{\rm p}_3^2
-\vec{\rm p}_4^2\right)
\end{eqnarray}
\end{mathletters}
where $H_{\rm S}$ is even
under charge conjugaison, and $H_{\rm A}$ odd. One
notices that $H$ and $H_{\rm S}$ have the same
dissociation  threshold, namely
$-(M^{-1}+m^{-1})^{-1}$ in our units. Now, $H_{\rm S}$
is stable by simple rescaling of the positronium-molecule case,
and the ground state of $H$ is lower than that of $H_{\rm S}$.
This latter  result comes from the
variational principle, with the ground state of $H_{\rm
S}$,  $\Psi(H_{\rm S})$, used as a trial wave function for $H$:
\begin{equation}
\label{HvsHS}
E(H)\le\langle\Psi(H_{\rm S})
|H_{\rm S}+H_{\rm A}|\Psi(H_{\rm S})\rangle
=E(H_{\rm S})
\end{equation}
since $\langle\Psi(H_{\rm S})|H_{\rm A}|\Psi(H_{\rm S})\rangle$
vanishes, by symmetry considerations.

It is amazing that every exotic molecule with the same structure as
hydrogen is stable. An example is  $D^+D^+\Omega^-\Omega^-$, with
charm $C=2$ and strangeness $S=-6$.

Previously, Abdel-Raouf \cite{Abdel} and Rebane \cite{Rebane}
stressed the regularity of the binding energy as a function of the
mass ratio $M/m$, but missed the fact that stability in the
positronium case implies stability for other
$(M^+M^+m^-m^-)$  configurations.

If one fixes for instance the scale by $m=1$, one can study
the lowest energy $E$ of $(M^+M^+m^-m^-)$ as a function of $s=1/M$,
or equivalently the fraction $x(s)$ defined as the positronium case
as
\begin{equation}
\label{General-x}
E(M^+M^+m^-m^-)=-{1+x(s)\over 1+s},
\end{equation}
$E_{\rm th}=-1/(1+s)$ being the threshold energy.
The fraction $x(s)$ takes the values $x_1=0.0303$ in the
positronium case $(s=1)$,  and $x_0=0.1745$ for $s=0$, the limit of
hydrogen  with infinitely massive protons. The values are taken from
the compilation by Rebane \cite{Rebane}. This means the
inequality (\ref{HvsHS}),
$E(H)<E(H_{\rm S})$, is actually observed.

Since $s$ enters the Hamiltonian linearly,  $E(s)$
is a concave function of $s$ \cite{Thir}. On can also combine convexity
and scaling  and deduce that $-1/E(s)$ is also concave, and this
provides a stronger  constraint \cite{Thir}.  The concavity of $-1/E$
leads to an upper bound on $x(s)$ for intermediate configurations
\begin{equation}
\label{limit-concavity}
1+x(s)\le {1+s\over2}\left[
{s\over1+x_1}+ {1-s\over2(1+x_0)}\right]^{-1}.
\end{equation}

\section{Asymmetric molecules}
\label{Aymmetric}
Let us go by steps toward $(m_1^+m_2^+m_3^-m_4^-)$,
which denotes the most general case. Consider first a different
arrangement of only two masses,
$(M^+m^+M^-m^-)$. We saw that it is stable for $M=m$. In case of a
large asymmetry $M\gg m$,  it seems to become unstable: the
hydrogen--antihydrogen system, for instance, hardly survives its decay
into $(p\bar p)+(e^+e^-)$.

A minimal domain of stability around $M=m$ can be derived from the
variational principle. One can fix the scale by setting
$M^{-1}=1-y$ and $m^{-1}=1+y$. Then
\begin{equation}
\label{H-y}
H(M^+m^+M^-m^-)={1\over2}
\left(\vec{\rm p}_1^2+\vec{\rm p}_2^2+\vec{\rm p}_3^2
+\vec{\rm p}_4^2\right)+V
+{y\over2}\left(-\vec{\rm p}_1^2+\vec{\rm p}_2^2-\vec{\rm p}_3^2
+\vec{\rm p}_4^2\right).
\end{equation}
The reasoning is the same as in Sec.~\ref{Hydrogen}.
The last term, antisymmetric under simultaneous $(1\leftrightarrow2)$
and $(3\leftrightarrow4)$  exchanges, lowers the
ground-state energy.  Then from Eq.~(\ref{Posit-x}),
\begin{equation}
\label{upper-y}
E(y)\le E_{\rm th}(0)(1+x_1).
\end{equation}
Meanwhile the threshold becomes
\begin{equation}
\label{Threshold-y}
E_{\rm th}(y)=-{1\over4(1-y)}-{1\over 4(1+y)}=
{E_{\rm th}(0)\over 1-y^2}.
\end{equation}
Thus stability remains at least as long as $y^2\le 1-(1+x_1)^{-1}$,
i.e.\ for                 \begin{equation}
\label{M-over-m}
0.70\le{M\over m}\le 1.43 \  .
\end{equation}
Now, if one starts from $(M^+m^+M^-m^-)$, and introduce
four different masses $m_i$ such that
\begin{equation}
\label{mass-distribution}
m_1\ge m_2,\quad m_3\ge m_4,\quad
m_1^{-1}+m_3^{-1}=2M^{-1},\quad m_2^{-1}+m_4^{-1}=2m^{-1},
\end{equation}
then one can rewrite the Hamiltonian as
\begin{eqnarray}
H(m_1^+m_2^+m_3^-m_4^-)&=&H(M^+m^+M^-m^-)\nonumber\\
& &+{1\over4}(m_1^{-1}-m_3^{-1})(\vec{\rm p}_1^2-
\vec{\rm p}_3^2) +{1\over 4}(m_2^{-1}-m_4^{-1})
(\vec{\rm p}_2^2-\vec{\rm p}_4^2),
\end{eqnarray}
and, again, the two last terms cannot do anything but lower
the ground-state energy, while the threshold does not change. Hence
stability is improved.

For instance $(\overline{\Omega^-}\Sigma^+\Xi^-\bar p)$ is
stable, as long as strong interactions are neglected, since the
constituent masses are here $1.67$, $1.20$, $1.32$ and $0.94\,{\rm
GeV}/c^2$, which fulfil (\ref{M-over-m}) and
(\ref{mass-distribution}). We are not too surprised to learn that  the
positronium hydride $(pe^+e^-e^-)$ is stable, by a  small margin
\cite{PosiHybride}: it corresponds to $M/m=2$, not too far outside our
minimal range  (\ref{M-over-m}),  and benefits from the large
asymmetry between particles 1 and 3.

The role of symmetries should be emphasised. Consider for
instance the stability of $(1^+,1^+,1^-,m^-)$, whose
threshold is
\begin{equation}
\label{th-example}
E_{\rm th}(m)=E_{\rm th}(1){1+\delta/4\over1+\delta/2}
\end{equation}
where $\delta=m^{-1}-1$ measures the departure from the symmetric case
of the positronium molecule.
One can first split the Hamiltonian into
\begin{equation}
\label{first-split-ex}
H(m)=H(1)+{\delta\over2}{\vec{\rm p}_4^2},
\end{equation}
and say that the ground state  should lie
 below the first order estimate.  The expectation
value of $\vec{\rm p}_4^2$ is easily estimated from the virial theorem
and the symmetries of the unperturbed wave function. Then
\begin{equation}
\label{first-bound-ex}
E(m)\le E(1)\left(1-{\delta\over4}\right),
\end{equation}
leading to a sufficient condition for stability
\begin{equation}
\label{first-dom-ex}
{1+\delta/4\over(1-\delta/4)(1+\delta/2)}=
1+{\delta^2\over8}+\cdots\le 1+x_1.
\end{equation}
If  one instead splits the Hamiltonian into pieces of well-identified
behaviour under permutations
\begin{equation}
\label{second-split-ex}
H=\left[\left(1+{\delta\over4}\right)\sum\vec{\rm p}_i^2 +V\right]
+{\delta\over8}\left(-\vec{\rm p}_1^2+\vec{\rm p}_2^2-
\vec{\rm p}_3^2+\vec{\rm p}_4^2\right)
+{\delta\over4}\left(-\vec{\rm p}_2^2+\vec{\rm p}_4^2\right),
\end{equation}
one gets
\begin{equation}
\label{second-bound-ex}
E(m)\le E(1){1\over1+\delta/4},
\end{equation}
corresponding to a wider range of stability
\begin{equation}
\label{second-dom-ex}
{ (1+\delta/4)^2\over1+\delta/2}=
1+{\delta^2\over 16}+\cdots\le 1+x_1.
\end{equation}

\section{Very asymmetric molecules}
\label{Very-asymmetric}

We already mentioned  that some configurations  are qualitatively
different from the positronium molecule, since their lowest
threshold  does not consist of two neutral atoms. These configurations
cannot studied by starting from $(e^+e^+e^-e^-)$, and implementing
more and more asymmetry.

Consider for instance the molecule
$(\Lambda_c^+\Lambda_c^+\Omega^-e^-)$,  with charmed hyperons, or any
similar case   where three particles are much heavier than the fourth
one, and form a stable ion. Its stability results from
the $(\Lambda_c^+\Lambda_c^+\Omega^-)$ ion lying below its
dissociation threshold into an atom and a charge,
as is the case for every $m_1^+m_2^+m_3^-$ ion with $m_1=m_2$
\cite{Hill,ArmB}. The electron is bound  around this ion,
but the electronic energy is negligible.

We expect interesting properties in the transition region where
a threshold consisting of 3-body ion and a charge becomes equal to
the lowest threshold made of two neutral atoms.
 The competing thresholds are likely to generate some
attraction in the whole 4-body system, and make it stable. This
particular situation  deserves some specific investigations.

\section{Tentative graphical summary}
\label{Graphics}
Thanks to the scaling properties of the Coulomb interaction, the study
of a system of $N$ given charges requires only $(N-1)$ independent
variables to scan all possible mass distributions.

In the case of 3 unit charges $q_i=\pm(-1,1,1)$, one could introduce
2 independent mass ratios. It was however found more convenient to use
3 barycentric coordinates $\alpha_i$, with $0\le\alpha_i\le1$, and a
normalisation $\sum\alpha_i=1$. Each case corresponds to a point
inside an equilateral triangle, in which a stability frontier
separates the stability from the instability areas. One can choose
the masses as barycentric coordinates, as in Fig.\ \ref{Figure1}.
The inverse masses of Fig.\ \ref{Figure2} prove more suited
for the mathematical analysis of
the observed convex behaviour of the frontier \cite{MRW}.

The generalisation to $N=4$ charges
corresponds the inner volume of a regular tetrahedron,
so that the stability frontier keeps the symmetries of the problem.
We shall summarise our results and our guesses
both with the masses $m_i$ as variables $(\sum m_i=1)$, or the inverse
masses $\alpha_i\propto m_i^{-1}$  $(\sum\alpha_i=1)$,
restricting ourselves to very schematic drawings of the unfolded
tetrahedrons. Future theoretical and numerical works will hopefully
make it possible to determine the frontier more accurately.

Consider first the representation in terms of the masses, in a
regular tetrahedron (ABCD) of unit height, so that $\sum m_i=1$, with
$m_1$ being the distance  to the face (BCD), etc. This corresponds
to Fig.\ \ref{Figure3}.

The summit A stands for the masses $m_i=(1,0,0,0)$, i.e.\ the limit
of a positronium hybride $(pe^+e^-e^-)$ with an infinitely massive
proton. It is stable, at least for $m_2=m_3=m_4$ \cite{PosiHybride}.
This means the stability domain reaches A, at least along the symmetry
axis. A similar situation is of course observed near B, C, and D.
To our knowledge, there is no result available on how
stability survives when the light masses $m_2$, $m_3$ and
$m_4$ are different, in particular for
$m_1\gg m_3\sim m_4\gg m_2$, corresponding to the trace of the
stability border on the face (ACD). For
$m_1\gg m_3=m_4\gg m_2$, we   have a heavy stable (1,3,4) ion
which attracts the charge $m_2$, a stable situation according to the
discussion in Sec.\ \ref{Very-asymmetric}.

The middle between A and C corresponds to masses $m_i\propto
(1,0,1,0)$, i.e.\ a configuration $(pe^+\bar pe^-)$,
 which seems unstable. We suspect there is no stability
at all along AC, and similarly along AD, BC and BD. This guess has of
course to be checked.

On the other hand, the middle between A and B describes the hydrogen
molecule $(ppe^-e^-)$ with $m_i=(1,1,0,0)$. This is a region of
confortable stability. We guess that stability holds all the way
along AB. There is no doubt  $(pp'e^-e^-)$ is stable as long as
both $m(p)\gg m(e)$ and $m(p')\gg m(e)$, since the Born--Oppenheimer
approximation holds in such case.
 Stability should remain, in our opinion,   for $m(p)\gg m(e)$ and
$m(p')\gtrsim m(e)$, as  long as the equality $m_3=m_4$ is
kept. This, indeed , suffices to make
the two threshods (1,3)+(2,4) and (1,4)+(2,3) degenerate, and
again, it is a general, though empirical, observation that competing
thresholds often generate attraction; one could also say that the
$(pe^-)$ and $(p'e^-)$ atoms have comparable Bohr radii, and their
overlap might lead to some favourable exchange forces between them.

The  face (BCD) corresponds to $m_1=0$. As discussed in
Sec. \ref{Very-asymmetric}, the stability domain is generally dictated
by the stability of the 3-body system $(m_2^+m_3^-m_4^-)$, which is
described in Fig. \ref{Figure1}. Exceptions are found near the edges
of (BCD), where we have more than one light particle. The situation
is of course identical on the other faces.

Consider now the tetrahedron (ABCD) with normalised inverse
distances, in Fig.~\ref{Figure4}. The summit A corresponds to the
$(e^+p\bar p\bar p)$, when one goes to A along the symmetry axis. This
is a stable protonium ion surronded by an electron, i.e.\ a
stable 4-body configuration. The shape of the stability
domain near A is more precisely seen on a cross-section (bcd) parallel
to (BCD) and close to A: the trace on (bcd) looks exactly as the
stability plot of Fig.\ \ref{Figure2}, because we are in the regime of
of Sec.\ \ref{Very-asymmetric} where one of the masses is much lighter
than of the others.

The sides AB and CD correspond to inverse masses
$(0,0,x,1-x)$, with $0\le x\le1$, i.e.\ to configurations $(ppee')$
with $m(e)\ll m(p)$ and $m(e')\ll m(p)$, but the ratio of $m(e)$ to
$m(e')$ arbitrary. We  believe they are stable.

 The sides AC, AD, BC, and BD have inverse masses of the type
$(x,0,1-x,0)$, i.e.\ $(e^+pe'\bar p)$,   for which dissociation seems
likely to occur spontaneously.

These speculations suggest the topology schematically
drawn in Fig.\ \ref{Figure4}. Note that with these variables, a face
such as (BCD) describes situations of the type
$\left( p e^+(e')^-(e'')^-\right)$, where (with obvious notations)
$m$, $m'$, and $m''$ are much lighter than $M$. It is stable for
$m=m'=m''$, with little excess binding \cite{PosiHybride}. We thus
expect the stability band to be rather narrow near the centre of
(BCD).

\section{Larger molecules}
\label{Larger-molec}
An obvious extension of the present study consists of
considering $N=N_++N_->4$ unit charges. The problem is in general
rather complex, due in particular to the proliferation of competing
thresholds, but one can retain from the study in the $N=4$ case that
simplifications occur when very different masses are involved.

Consider for instance $N=5$, with masses and charges
$(M^+M^+M^-m^-m^-)$, and $M\gg m$. The heavy core $(M^+M^+M^-)$
is stable, by simple rescaling of the positronium-ion case
$(e^+e^-e^-)$. It acts as a localised positive charge, and  should
bind two electrons, with a wave function very similar to that of the
familiar H$^-(pe^-e^-)$ ion. a similar reasoning was applied to
molecular complexes involving muons and electrons \cite{Faifman}. If
$(M^+M^+M^-m^-m^-)$ is stable for large $M/m$, a cooled antiproton
could be trapped in a hydrogen molecule, before annihilating.
According to the current analysis \cite{Klempt}, when a $\bar p$ is
captured in hydrogen, it quickly expels the electrons by Auger
emission;  a protonium atom $(p\bar p)$ is formed in a highly excited
state, and it rapidly decays toward states of low orbital momemtum,
where its annihilates. Recently some events with delayed annihilation
have been reported, stimulating some theoretical studies
\cite{Yamazaki,Richter}: some metastable states with an antiproton, a
nucleus and some electrons might be formed in rare occasions. The
above $(pp\bar pe^-e^-)$ compound is another new possibility.

Similarly, for $N=6$, and $N_+=N_-=3$, the state
$(ppe^+\bar pe^-e^-)$ is probably stable, due to the combined
stability of the $(pp\bar p)$ ion and of the positronium hybride
$(pe^+e^-e^-)$.

The symmetric case of $N=6$, $(m^+m^+m^+m^-m^-m^-)$, is much more
delicate. Stability seems excluded for actual electrons, due to the
Pauli principle, but one can address the question for pions, $\pi^+$
and $\pi^-$, which are bosons. The lowest threshold is
$(m^+m^+m^-m^-)+(m^+m^-)$, so one has to show that the lowest energy
$E$ decreases faster than $-N$, already for small $N$. For large $N$,
a behaviour $E\propto -N^{7/5}$ has been identified \cite{Conlon}.

\section{Applications to hadron spectroscopy}
\label{sec-tetraquark}
A simple, but phenomenologically successful, description of the
hadron spectrum is the non-relativistic quark model. Mesons are
quark--antiquark bound states, and baryons consist of three quarks.
An interesting
property is {\sl flavour independence}. At first
approximation, i.e.\ without spin forces and relativistic
corrections, the potential is the same whatever
quarks,  $u$, $d$, $s$, $c$, or $b$, are
bound together. This is reminiscent from
 atomic physics, where the very same $-r^{-1}$ potential acts in
hydrogen, positronium, protonium, etc.

In quark models with flavour independence, one expects
some of patterns observed in atomic physics, those which are
 due to the universality of the potential, and independent of its
particular Coulombic shape. For instance, the
convexity property mentioned in Sec.~\ref{Stability} for the 2-body
energies is satisfied, and  the
inequality
\begin{equation}\label{Tetraquark} Q\overline{Q}+q\bar q\le2Q\bar q
\end{equation}
has indeed been noticed \cite{BerMar}, and observed in the experimental
spectrum \cite{PDG}.

We have seen that among the 4-body molecules, $(ppe^-e^-)$ was the most
stable. It is not surprising that explicit four-quark calculations
 \cite{BadHonnef,TetraQuark} have
led to the prediction that the exotic configuration ($QQ\bar q\bar q$)
becomes stable if the mass ratio $m(Q)/m(q)$ is large
enough, whereas the equal-mass case
$QQ\overline{Q}\overline{Q}$ seems unbound.
In this field,
one call ``exotic'' a state whose quantum numbers, in
particular flavour, cannot be matched by ordinary
quark-antiquark or three-quark structures. Other
multiquarks are predicted on the basis   of spin-dependent
forces, but this is out of the scope of this paper.  It is
hoped that double-charm, or charm-and-beauty
spectroscopy will reveal some good
surprises.

\section{Conclusions}
\label{Conclusions}

Let us summarise our results  about 4
unit-charge systems, which bear  a sharply decreasing degree of
rigour.

We first recalled the beautiful proof by Hylleraas and Ore of the
stability of the positronium molecule \cite{Ore}. It can be
supplemented by a rigourous study of the ordering of the various
possible thresholds \cite{Froehlich,Seifert}.

Once the stability of $(e^+e^+e^-e^-)$ is established, that of the
hydrogen-like molecules $(M^+M^+m^-m^-)$ follows rigourously
\cite{BadHonnef,Froehlich}. This is essentially a consequence of the
variational principle.

Now, one can prove a minimal extension of the stability domain
toward other directions like $(M^+m^+M^-m^-)$ or
$(m_1^+m_2^+m_3^-m_4^-)$, if one takes for granted a non-rigourous
input, namely the actual binding energy of $(e^+e^+e^-e^-)$, which
results from numerical variational calculations.

The above results are far from covering all possible configurations.
To picture how the stability domain could look like,
we have substituted some missing results by simple guesses, which
have to be checked. For instance, we feel that the asymmetric
hydrogen molecule $(pp'e^-e^-)$ remain stable whatever mass is given
to $p'$, as long as $m(p)\gg m(e)$. We have also assumed without proof
that $(pe^+\bar p'e^-)$ immediately breaks into $(p\bar
p')+(e^+e^-)$, except near the limit $m(p')\rightarrow m(e)$.

We also presented some plausible but non rigourous extensions to
larger molecules and to multiquark systems in the quark model.

In spite of its weaknesses, this study illustrates once more
how quantum binding energies  have a simple and
well-controlled  behaviour when one varies the constituent masses
without changing the  potential. It is hoped it will stimulate some
interest for the investigations which remain to be done.

\acknowledgments

I would like to thank D.M.\ Brink  for useful comments
and discussions. The hospitality provided by ECT* at Trento
is gratefully acknowledged. The Institut des Sciences Nucl\'eaires
is supported by Universit\'e Joseph Fourier and CNRS-IN2P3.


\begin{figure}
\setbox1=\vtop{\vglue 8cm
\vglue .5cm}
\setbox2=\vtop{\vglue 8cm
\vglue .5cm}
\hbox{\hskip 1.0cm\raisebox{2.0cm}{\box1}
\hskip 8.5cm\raisebox{2.0cm}{\box2}}
\caption{\label{Figure1} Graphical representation of the
stability domain for 3 unit charges \protect{$q_i=\pm(-1,1,1)$}
with  normalised masses \protect{$m_i$}, \protect{$\sum m_i=1$}.
The stability domain includes the symmetry axis (dotted line)
where \protect{$m_2=m_3$}.}
\end{figure}

\begin{figure}
\setbox3=\vtop{\vglue 8.0cm
\vglue .5cm}
\setbox4=\vtop{\vglue 8.0cm
\vglue .5cm}
\hbox{\hskip 1.0cm\raisebox{2.0cm}{\box3}
\hskip 8.5cm\raisebox{2.0cm}{\box4}}
\caption{\label{Figure2} Graphical representation of the
stability domain for 3 unit charges \protect{$q_i=\pm(-1,1,1)$}
with  normalised inverse masses \protect{$\alpha_i=1/m_i$},
\protect{$\sum \alpha_i=1$}.
The stability domain includes the symmetry axis (dotted line)
where \protect{$\alpha_2=\alpha_3$}.}
\end{figure}

\begin{figure}
\setbox5=\vtop{\vglue 12cm
\vglue .5cm}
\hbox{\hskip 3.0cm\raisebox{1.0cm}{\box5}}
\caption{\label{Figure3} Schematic picture of the
 stability domain for 4 unit charges
\protect{$q_i=(-1,-1,1,1)$}. The  stability  frontier
is shown by its trace on the unfolded faces of the tetrahedron (ABCD)
of normalised  masses  \protect{$m_i$}, \protect{$\sum m_i=1$}.}
\end{figure}

\begin{figure}
\setbox6=\vtop{\vglue 12cm
\vglue .5cm}
\hbox{\hskip 3.0cm\raisebox{1.0cm}{\box6}}
\caption{\label{Figure4} Schematic picture of the
 stability domain for 4 unit charges
\protect{$q_i=(-1,-1,1,1)$}. The  stability  frontier
is shown by its trace on the unfolded faces of the tetrahedron (ABCD)
of inverse  masses  \protect{$\alpha_i=1/m_i$}, with
normalisation \protect{$\sum \alpha_i=1$}.
The trace of the stability domain in the
section (bcd) below the summit A looks like the 3-body domain
of Fig.\ \protect{\ref{Figure2}}. }
\end{figure}

\end {document}